# Declarative Reconfigurable Trust Management

William R. Marczak[1]   David Zook[2]   Wenchao Zhou[1]
Molham Aref[2]   Boon Thau Loo[1]

[1] Univ. of Pennsylvania, {wrm,wenchaoz,boonloo}@seas.upenn.edu
[2] LogicBlox, {david.zook,molham.aref}@LogicBlox.com

## ABSTRACT

In recent years, there has been a proliferation of declarative logic-based trust management languages and systems proposed to ease the description, configuration, and enforcement of security policies. These systems have different tradeoffs in expressiveness and complexity, depending on the security constructs (e.g. authentication, delegation, secrecy, etc.) that are supported, and the assumed trust level and scale of the execution environment. In this paper, we present *LBTrust*, a unified declarative system for *reconfigurable trust management*, where various security constructs can be customized and composed in a declarative fashion. We present an initial proof-of-concept implementation of *LBTrust* using LogicBlox, an emerging commercial Datalog-based platform for enterprise software systems. The LogicBlox language enhances Datalog in a variety of ways, including *constraints* and *meta-programming*, as well as support for programmer-defined constraints on the meta-model itself – *meta-constraints* – which act to restrict the set of allowable programs. *LBTrust* utilizes LogicBlox's meta-programming and meta-constraints to enable customizable cryptographic, partitioning and distribution strategies based on the execution environment. We present use cases of *LBTrust* based on three trust management systems (Binder, D1LP, and Secure Network Datalog), and provide a preliminary evaluation of a Binder-based trust management system.

## 1. INTRODUCTION

*Trust management* is an essential and important component of security and it is pervasive in computer systems. Trust management is broadly defined as the process of formulating access control policies and security credentials, determining whether particular sets of credentials satisfy the relevant policies, and deferring trust to third parties [8]. Over the years, logical ideas and tools have been used to explain and improve trust management, particularly to implement access control in a multi-user distributed environment. Several declarative logic-based languages (e.g. Binder [12], Cassandra [7], D1LP [15], SD3 [13]) have been proposed to ease the process of expressing, analyzing, and encoding security policies.

These proposals have different tradeoffs in expressiveness and complexity, depending on the security constructs (e.g. authentication, delegation, speaks-for, etc.) that are supported, as well as the trust level and distribution of their assumed environments. At the minimum, Binder (one of the simplest languages) supports the logical `says` operator for *authentication*, where principals that assert facts must first validate their identity in a secure fashion.

Other trust management systems have explored additional security constructs for *secrecy* [3] and *encrypted facts* [17], ensuring that only authorized principals can interpret facts in distributed settings. In order to differentiate principals based on their capabilities, one can further incorporate the notions of *speaks-for* [14] and *restricted delegation* [15] into the language, which allow principals to delegate the responsibility of selected policy decisions to other principals.

All in all, there is an inherent tradeoff between expressiveness and complexity across these spectrum of languages, and while each language may be intriguing in isolation, one would not want to combine them all indiscriminately. Interestingly, it has been shown previously [2, 4] that distributed trust management languages share similarities to data integration languages (e.g. Tsimmis [10] as observed by Abadi *et al.* [2]) and the distributed Datalog languages proposed for declarative networking [9], by supporting the notion of *context* (location) to identify *components* (nodes) in distributed systems, indicating that ideas and methods from the database community may be applicable to processing security policies. These similarities provide the possibility of unifying these declarative languages to create an integrated system.

In this paper, we present *LBTrust*, a unified declarative system for *reconfigurable trust management*, where various logical security constructs can be customized and composed using a variant of the Datalog language. We present an initial proof-of-concept implementation of *LBTrust* using *LogicBlox* [16], an emerging commercial Datalog-based platform for enterprise software systems. The LogicBlox language provides several enhancements to Datalog, including *constraints* and *meta-programming*. LogicBlox's meta-programming is based on a *meta-model* similar to the recently proposed Evita Raced [11]), which features a bootstrapped meta-circular compiler implemented in Datalog. Unlike Evita Raced, LogicBlox also supports applying programmer-defined constraints to the meta-model itself – *meta-constraints* – which act to restrict the allowable programs. *LBTrust* utilizes LogicBlox's meta-programming and meta-constraints to enable customizable cryptographic, partitioning and distribution strategies based on the deployed environment.

Using *LBTrust*, we first demonstrate that a variety of security primitives for authentication, confidentiality, integrity, speaks-for, and restricted delegation (used in D1LP) can be supported. Based on these primitives, we present a detailed use case of *LBTrust* to support the Binder [12] trust management system. We further demonstrate the use of *LBTrust* to implement *Secure Network Datalog* [4, 19], a secure declarative networking language that unifies Binder and with the *Network Datalog* language used in declarative networking. Finally, we show an initial evaluation of the performance overhead of *LBTrust* based on *LBTrust*'s implementation of Binder.

Because it is a unified declarative platform, *LBTrust* provides a basis for comparison across different trust management systems, and potentially provides avenues for better analyzing security properties across these various languages.

## 2. BACKGROUND

### 2.1 Datalog

We first provide a short review of Datalog, following the conventions in Ramakrishnan and Ullman's survey [18]. A Datalog program consists of a set of declarative *rules*. Each rule has the form `p <- q1, q2, ..., qn.`, which can be read informally as "q1 and q2 and ... and qn implies p". Here, p is the *head* of the rule, and q1, q2, ..., qn is a list of *literals* that constitutes the *body* of the rule. A literal is a possibly negated atom. An atom is a *predicate* applied to a list of *terms*, each of which is either a *constant* or a *variable*. The names of predicates, function symbols, and constants begin with a lowercase letter, while variable names begin with an uppercase letter. Negation may not occur in the head of a rule, and in the body it must be *safe* – every variable occurring in a negated literal must also occur somewhere in a non-negated literal. Also, for readability, *solitary* variables – those which occur just once in a rule – are often replaced with an underscore (`_`).

Each predicate occurring in the head of a rule is called *intensional*, while all the other predicates are called *extensional*. A Datalog program takes as input a model (an assignment of values) of the extensional predicates and derives a minimal model of the intensional predicates consistent with the logical meaning of the rules.

LogicBlox (and most Datalog implementations) provide built-in functions for equality, arithmetic, and aggregation (totaling and counting), as well as built-in predicates for common types (numbers and strings). Also, it easily can be shown that an arbitrary nesting of negation, conjunction, and disjunction may be used in the body of a rule. Such a rule may be translated into strict Datalog rules by (1) translating the body into Disjunctive Normal Form (DNF), and (2) splitting the original rule into a separate rule for each resulting alternative, duplicating the original head. We use a left-arrow (`<-`) for logical implication, a comma (`,`) for conjunction, a semicolon (`;`) for disjunction, a bang (`!`) for negation, and parentheses for grouping.

### 2.2 Logic-based Trust Management

To illustrate logic-based trust management languages [1], we provide an example language based on Binder [12]. Binder expresses access control policies in a multi-user distributed environment. A Binder program is a set of Datalog-style logical rules. In addition, Binder has the notion of a *context* – a component in a distributed environment – and a distinguished operator called `says`. For instance, in Binder we can write:

```
b1: access(P,O,read) <- good(P).
b2: access(P,O,read) <- bob says access(P,O,read).
```

The `says` operator implements a common logical construct in authentication [14], where we assert "p says s" if the principal p supports the statement s. The above rules b1 and b2 can be read as "any principal P may access any object O in read mode if P is good or if bob says that P may do so". The `says` operator abstracts from the details of authentication.

A principal in Binder refers to a component in a distributed environment. Each principal has its own local *context* where its rules reside. Binder assumes an *untrusted* network, where different components can serve different roles, running distinct sets of rules. Because of the lack of trust among nodes, a component does not have control over rule execution at other nodes. Instead, Binder allows separate programs to interoperate correctly and securely via the export and import of rules and derived tuples across contexts. For example, rule b2 can be a local rule that is executing in the context of principal alice, which imports derived access tuples from the principal bob into its local context via `bob says access(p,o,read)` in its rule body.

Binder specifies an asymmetric key signature scheme, such as RSA, for the "says" construct. In a hostile world, "says" may require this, but in a more benign world, one may wish to trade some security for efficiency, and configure "says" to simply append cleartext principal headers to messages. Somewhere in between, the use of cryptographic signatures may be applied only to certain important messages, or when communicating with specific principals. Binder does not provide any leverage in deciding how this tradeoff should be made.

In addition to constructs for authentication, declarative trust management systems often feature security constructs for *integrity*, *secrecy*, and *delegation*. The D1LP [15] language further supports constructs that implement distributed vote-based agreement, where a fact in the rule head is derived only when *k-out-of-n principals* in a rule body predicate derive a similar fact concurrently. We will revisit *LBTrust*'s support for these security constructs in Section 4.

## 3. LB-TRUST ARCHITECTURE

*LBTrust* is implemented using *LogicBlox*, a commercial platform for building enterprise-scale corporate planning and pricing applications, which feature analyses requiring aggregation across very large data sets, combined with simulation and modeling techniques.

LogicBlox contains a Datalog-based logic programming language enhanced with a variety of features, including functional dependencies, aggregation functions, schema constraints, static type-checking, tuple-generation, temporal logic support, predicate partitioning, distributed computation, and meta-programming. LogicBlox further allows application-defined libraries of custom predicates to be imported, such as the cryptographic functions required for implementing certain security constructs. Also essential for these constructs is LogicBlox's support for applying schema constraints to the meta-model, which we call *meta-constraints*. To set the stage for presenting *LBTrust*'s implementation of various security constructs, this section explains LogicBlox's facilities for supporting schema constraints, meta-programming, and distributed computation.

### 3.1 Execution Environment

LogicBlox utilizes a bottom-up semi-naïve fixpoint [18] execution model for executing Datalog programs. LogicBlox provides a query interface for submitting a program for compilation and execution within a *workspace*. A *workspace* in LogicBlox is essentially a database instance which contains a set of predicate definitions[1] and a set of *active* rules (similar to *continuous queries*). Within a designated workspace, the LogicBlox API allows an application to query and modify the data defined by the workspace, including adding/removing facts and rules. When predicate data is modified, the active rules are incrementally recomputed.

### 3.2 Constraints

Unlike a rule, which calculates new values for a predicate, a *schema constraint* – such as a referential integrity constraint [5]) – *restricts* a predicate's allowed values. LogicBlox adds schema constraints to Datalog by means of the special predicate `fail()`. If any rule defines `fail()` to be true, then the evaluation of the Datalog program fails by terminating with an error.

For example, a schema constraint for the Binder program given above might require that any value occurring in the first argument of the `access` predicate also occur in the `principal` predicate. This constraint can be expressed as a rule:

```
fail() <- access(P,O,M), !principal(P).
```

This rule defines `fail()` to be true if, for any assignment of values to the variables P, O, and M, the atom `access(P,O,M)` is true but `principal(P)` is false.

Constraints expressed using `fail()` can often be un-intuitive. So as a notational convenience LogicBlox supports a *positive* form for constraints, indicated with a right arrow (->). If $F_1$ and $F_2$ are arbitrary nestings of conjunction, disjunction and negation, then the logical meaning of $F_1$ -> $F_2$. is `fail()` <- $F_1$, !($F_2$). For example, the positive form of the constraint given above is:

```
access(P,O,M) -> principal(P).
```

Typically every argument of `access` would be constrained:

```
access(P,O,M) -> principal(P), object(O), mode(M).
```

Informally, this may be read as "for any values of P, O and M, whenever `access(P,O,M)`, then require `principal(P)` and `object(O)` and `mode(M)`". In fact, in LogicBlox, a *type* is considered to be a unary predicate (representing a set of values). Hence, in LogicBlox, this kind of schema constraint acts as a *type declaration*. The use of types and type-checking (statically, and dynamically when rules are added to workspaces) ensures that only type-safe LogicBlox programs are executed.

---

[1]A predicate definition declares both the logical attributes of a predicate, such as its name and arity; and also the physical attributes for the purposes of cost-based optimizations, such as the predicate's data storage format, data location, and population density statistics.

### 3.3 Meta-Programming

The basis of the LogicBlox meta-programming feature is a special set of predicates called the *meta-model*, whose (implicit) type declarations expressed as constraints are shown in Figure 1. Each LogicBlox workspace stores an `active` table that contains all the rule identifiers of active rules within the workspace. When a rule R is added to the workspace's active rules, it is translated into a set of facts (e.g. `rule`, `head`, `body`, etc.) in the meta-model, and its rule id is stored in the `active` table.

```
rule(R) ->.
head(R,A)-> rule(R), atom(A).
body(R,A)-> rule(R), atom(A).

atom(A)-> .
functor(A,P)-> atom(A), predicate(P).
arg(A,I,T)-> atom(A), int(I), term(T).
negated(A)-> atom(A).

term(T)-> .
variable(X)-> term(X).
vname(X,N)-> variable(X), string(N).
constant(C)-> term(C).
value(C,V)-> constant(C), string(V).

predicate(P)-> .
pname(P,N)-> predicate(P), string(N).
```

**Figure 1: The Meta-Model**

The significance of the meta-model is that programmer-defined rules may refer to the meta-model. For example, an active rule may perform *reflection* (i.e. query for the program's structure) by referring to meta-model predicates in its body. Or, a rule may perform *code generation* (adding or rewriting existing rules) by referring to the meta-model in its head. If the evaluation of a rule puts new facts into the meta-model, then those new facts turn into a *new rule* which must itself be evaluated.

A special case of reflection is a schema constraint that refers to meta-model predicates (a *meta-constraint*). While meta-constraints are usefully generally for imposing integrity constraints similar to those in databases, they are particularly useful in the context of *LBTrust* for expressing security restrictions. To illustrate, assume we wanted to require that a principal may only read predicates to which they have been granted access. Datalog (without constraints) provides no way to enforce this requirement, because there is no way to prohibit an attempted access.

To support this restriction in *LBTrust*, we first define an `owner` predicate that associates a rule with the principal that added that rule, and an `access` predicate that represents access rights to a predicate. Note that we leverage `predicate`, a meta-model predicate that contains a unique entry for each predicate defined in the workspace (including `predicate`).

```
owner(R,P) -> rule(R), principal(P).
access(U,P,M) ->
   principal(U), predicate(P), mode(M).
```

With our schema defined, we then apply the following meta-constraint, which provides our desired prohibition. It says that for any principal `U` who owns a rule with predicate `P` in the body, there must be a fact in the `access` predicate granting `U` the right to read `P`.

```
owner(U, [| A <- P(T2*), A*. |]) ->
   access(U,P,read).
```

The above example illustrates that meta-programming is facilitated in LogicBlox by the introduction of the *quoted code* term – a rule or atom surrounded by the *code-quotes*: `[|` and `|]`. Inside the code-quotes is a *code pattern* that matches one or more rules. The star (*) represents the Kleene star – a repetition of the pattern preceding it. The capital letters in the pattern are *meta-variables* – variables that represent pieces of code, and can occur in non-standard places, like functors and atoms. The types of the meta-variables are determined by their position in the pattern. The LogicBlox compiler translates the code inside the code-quotes into a conjunction of atoms on the meta-model representing the quoted code. For example, the above meta-constraint translates into:

```
owner(U,R1), rule(R1), body(R1,A1),
atom(A1), functor(A1,P) ->
   access(U,P,read).
```

The variables `R1` and `A1` are freshly generated by the translation. Note that the meta-variable `P` occurs outside the quoted code – its value is *unquoted* in-place into the pattern, without any special unquoting operator.

### 3.4 Partitioning

Partitioning is a mechanism for logically separating facts based on their attributes. In trust management systems, data is partitioned by principal, with each partition typically called a principal's *context*. A context stores all facts local to a principal, and communication between contexts represents communication between principals.

In order to support partitioning, the `predicate` predicate in the meta-model is *higher-order*, meaning that it takes a predicate as an argument. In general, consider a predicate `p` with $n$ arguments where `t1(X1)` denotes the type of `X1`:

```
p(X1, ..., Xn) -> t1(X1), ..., tn(Xn).
```

To partition `p` based on the first attribute `X1`, one can rewrite the above as follows:

```
p'[X1](X2, ..., Xn) -> t1(X1), ..., tn(Xn).
```

`p'` is a higher-order predicate, where for a given `x1`, `p'[xi]` refers to all `p` predicates whose first argument has value `x1`.[2]

The partitioning of `p` does not change the set of data that can be stored in `p`, but instead *partitions* the data into subsets (the `p'` predicates). A regular Datalog rule of the form `p'[X1](X2, ..., Xn) <- p(X1, ..., Xn)` can be used to initialize `p'` partitions based on the input table `p`.

---
[2]This general rewrite technique for partitioning is generally known as *currying* in the functional programming world, a generic logical transformation that uses one of the arguments as the partitioning argument.

### 3.5 Distribution

In a distributed setting, principals may reside on different nodes, and the execution of security policies may result in an exchange of rules, similar to Binder's transfer of rules across contexts. In LogicBlox, logical partitioning and distribution are separated, hence providing *location transparency*, where a multi-principal security policy can be distributed in a customized fashion based on the deployed execution environment (e.g. single vs multiple principals per physical host).

To support distribution, LogicBlox introduces a special meta-model predicate `predNode` which is used to customize the location of a predicate `P`:

```
predNode(P,N) -> predicate(P), node(N).
```

So if the application were to associate each value of `X1` with a node using some predicate:

```
locX1(X1,N) -> t1(X1), node(N).
```

Then the application could distribute the subsets of `p` using the *placement rule*:

```
predNode(p'[X1],N) <- locX1(X1,N).
```

This rule takes each value of `X1`, looks up its associated network node `N`, and then places the corresponding subset of `p'` on `N`. Using these techniques, which combine meta-model predicates with *currying*, a LogicBlox application can use ordinary Datalog rules to partition and distribute the data in its predicates.

## 4. SECURITY PRIMITIVES IN LBTRUST

In this section, we demonstrate how various security primitives can be customized and supported by *LBTrust*. This is by no means intended to be an exhaustive coverage of the possibilities. Our main goal here is to illustrate the key language features of *LBTrust*, and highlight the flexibility and compactness of *LBTrust* in supporting various security primitives. We will build upon these primitives in our next section when we present case studies of languages enabled by *LBTrust*.

### 4.1 Authentication

*Authentication* is a central component of security, where the identity of a principal is established and verified. Authentication is essential for *authorization*, where an authenticated principal is granted access to perform actions on shared resources. Practically all logic-based trust management languages provide some language support for expressing authentication, typically via the `says` operator described in Section 2, which associates a principal with a statement. In each case, the semantics of `says` are hardwired into the system as an add-on to the logic programming language. In *LBTrust*, however, the `says` concept is configured in the same language as the policy, using features not specifically designed to support security concerns – except for cryptographic primitives used to implement various authentication schemes.

The simplest way to associate a principal with every fact in a predicate `P(T*)` is to add an extra argument representing the principal who said the fact: `P(U,T*)`. In *LBTrust*, we represent `says(U1,U2,R)` as a meta-predicate that associates a Datalog rule (R) with both the source principal who said the rule (U1), and the destination principal to whom the rule is said (U2). Note that while communication occurs in the form of rules, we can also communicate facts (rules with an empty body). We define the `says` predicate below.

```
says0: says(U1,U2,R) -> prin(U1), prin(U2), rule(R).
says1: active(R) <- says(_,me,R).
```

Rule `says0` is a type declaration for `says`. The `says1` rule automatically moves any rule R said to the local principal (indicated by the `me` keyword), into the built-in meta-predicate `active`, which automatically activates R.

Using the `says` predicate, authorization may be implemented with some simple meta-constraints. The following constraints restrict read and write access to predicates respectively.

```
says(U,me [| A <- P(T*), A*. |]) -> mayRead(U,P).
says(U,me [| P(T*) <- A*. |]) -> mayWrite(U,P).
```

### 4.1.1 Authenticated Communication

To enable communication between principals, we introduce the `export` predicate as well as additional meta-rules and meta-constraints. The following rules implement rule export using the RSA authentication scheme (`rsasign` and `rsaverify`).

```
exp0: export[U1](U2,R,S) -> prin(U1), prin(U2),
      rule(R), string(S).
exp1: export[U2](me,R,S) <- says(me,U2,R),
      rsasign(R,S,K), rsaprivkey(me,K).
```

Consider a local principal `me` that wishes to export rule R to another principal U2 via the `says` predicate. Rule `exp0` declares the type definitions of the `export` predicate that will be used for exporting the rule R with its signature to the destination principal U2. `export` has a placement policy that assigns the location of each partition to match the location of the corresponding principal.

Rule `exp1` calculates the appropriate RSA signature S using the private key of the local principal, and copies the rule into the destination principal's partition of the `export` predicate. The following rules would then run at the destination principal (U2), to import received rules. Note that in this case, `me` refers to principal U2.

```
exp2: says(U,me,R) <- export[me](U,R,S).
exp3: says(U,me,R) -> export[me](U,R,S),
      rsapubkey(U,K), rsaverify(R,S,K).
```

Rule `exp2` copies the received rule from the `export` predicate into the local `says` predicate. Finally, `exp3` verifies the signature of the new rule using the source principal's public key.

### 4.1.2 Alternative Authentication Schemes

Because the signature generation and verification has been defined in a Datalog rule, it is easy to replace the RSA scheme above with an alternate scheme. To illustrate, we demonstrate signing each message with a *message authentication code* (MAC), typically a 160-bit SHA-1 cryptographic hash of the message data and a secret key shared between the two communicating principals. The choice of an alternative signature generation scheme is often a tradeoff between security and performance. For example, MAC is computationally less expensive but requires the use of shared symmetric keys among principals that wish to communicate. Also, the use of a symmetric key to generate a rule signature implies that the signature will only be verifiable at principals that have the same key as the signer. The following rules implement an HMAC-SHA1 signature scheme.

```
exp1': export[U2](me,R,S) <- says(me,U2,R),
       hmacsign(R,K,S), sharedsecret(me,U2,K).
exp3': says(U,me,R) -> export[me](U,R,S),
       sharedsecret(me,U,K), hmacverify(R,S,K).
```

Interestingly, only two rules (`exp1'` and `exp3'`) need to be modified, while the trust policies that utilize the `says` predicate remain unchanged, demonstrating the ease with which new authentication schemes can be enabled by *LBTrust*.

### 4.1.3 Confidentiality and Integrity

*LBTrust* can support *confidentiality*, ensuring rules cannot be interpreted by unauthorized principals in a distributed setting, and *integrity*, ensuring data is not corrupted. This requires the addition of built-in predicates representing various encryption and integrity schemes such as checksums and cryptographic hashes.

## 4.2 Delegation

Often-times in security, it is useful to establish a *chain of trust* among different principals. This is particularly useful for performing *delegation*, where different principals may choose to assign capabilities to other principals, either for performance, accessibility, or security reasons. For example, a principal may delegate the authority to associate principals with public keys to a certificate authority. Or a credit card issuer may wish to delegate authority to a credit rating agency to associate a credit score with an individual.

An early version of delegation is the *speaks-for* construct. Adopting the definition based on Lampson's security survey paper [14], *speaks-for* works as follows. Consider a statement of the form "principal U1 speaks for principal U2." The logical meaning behind "speaks for" is that if U1 says something, then U2 says it too. So for example, `alice` can say that `bob` speaks for her by activating the following meta-rule, which activates any rule R said by `bob`.

```
sf0: active(R) <- says(bob,me,R).
```

A speaks-for rule is a special case of delegation where a principal delegates all authority to another principal. In practice, it is useful to restrict this delegation to a specific predicate. To ease the specification of complex delegation policies, *LBTrust* defines a `delegates` predicate whose type declaration is shown in `del0`. In the example below, `delegates(U1,U2,P)` denotes that U1 delegates the responsibility of deriving predicate P to U2 – essentially expressing the *speaks-for* construct where U2 speaks for U1 with respect to P.

```
del0: delegates(U1,U2,P) -> prin(U1), prin(U2),
      predicate(P).
del1: active([| active(R) <- says(U2,me,R),
      R = [| P(T*) <- A*. |]. |]) <-
      delegates(me,U2,p).
```

The `del1` meta-rule defines the `says` predicate in terms of the `delegates` predicate: whenever a delegation fact is added, the appropriate speak-for rule is automatically generated.

### 4.2.1 Delegation Depth and Width

Sometimes it is useful to restrict delegation authority [15]. For instance, we may restrict one or both of the *depth* – the maximum permitted length of the delegation chain – and the *width* – the set of principals allowed to be part of the chain.

The following meta-rules declare and enforce a `delDepth` predicate. The `inferredDelDepth` predicate takes the originally specified delegation depths and infers new depth restrictions. The base case is when a principal U1 delegates to U2 with depth limitation N=0. In this case, as expressed by meta-constraint dd4, any delegation by U2 conflicts with the limitation. The recursive case, as expressed by the rule defining `inferredDelDepth`, is when U1 delegates to U2 with a depth limitation of N > 0. In this case, if U2 delegates to some other principal U3, then a new limit of N-1 is inferred between U2 and U3. Similar meta-rules can be formulated to enforce delegation width restrictions.

```
dd0: delDepth(U1,U2,P,N) -> prin(U1), prin(U2),
        predicate(P), int[64](N).
dd1: inferredDelDepth(U1,U2,P,N) -> prin(U1),
        prin(U2), predicate(P), int[64](N).
dd2: inferredDelDepth(me,U,P,N) <-
        delDepth(me,U,P,N).
dd3: says(me,U,[|
    inferredDelDepth(me,U,P,N-1). |]) <-
        inferredDelDepth(me,U,P,N),
        delegates(me,U,P), N>0.
dd4: inferredDelDepth(_,me,P,0) ->
        !delegates(me,_,P).
```

An interesting case arises if a non-conforming delegation exists before a delegation depth restriction is added. The rule dd3 will propagate an inferred delegation depth of 0 to the principal with the non-conforming delegation, causing a violation of the dd4 constraint. However, none of the principals in the delegation chain up to that point will be aware of the violation.

### 4.2.2 Delegation Thresholds

Another delegation variant is the use of threshold structures. An unweighted threshold structure will authorize some operation if any k out of n principals concur. For example, a bank may consider a customer's credit okay if at least three credit bureaus do. This is easily expressed in *LBTrust* using the `count` aggregation:

```
wd0: creditOK(C) -> customer(C).
wd1: creditOK(C) <-
        creditOKCount(C,N), N >= 3.
wd2: creditOKCount(C,N) <- agg<<N = count(U)>>
        pringroup(U,creditBureau),
        says(U,me, [| creditOK(C). |]).
```

It is also straightforward to generalize these rules to handle more sophisticated threshold structures such as *weighted delegation*, where different credit bureaus have different reliability factors assigned to them. Rule wd2 above would be modified to use the `total` aggregation.

## 5. CASE STUDIES

In this section, we focus on how *LBTrust* can leverage the basic security constructs presented in the previous section to implement trust management languages. We focus on two case studies: Binder and the Secure Network Datalog (*SeNDlog*) language used in declarative networking.

### 5.1 Binder

As described in Section 2.2, Binder is a logic-based trust management system, which extends Datalog with the `says` construct and the notion of communication across *contexts*. Each component, or principal in the distributed system has a local Binder context. Binder contexts correspond to LogicBlox *workspaces* described in Section 3. To authenticate facts asserted by principals, Binder uses certificates signed with the private key of the sending principal. Certificates are imported by prefixing the `says` operator with a public key representing the context to import from. In our implementation of Binder, we use the `says` predicate defined in Section 4.

To illustrate, the *LBTrust* equivalent to the Binder rule b2 presented in Section 2.2 is:

```
bex1': access(P,O,read) <-
        says(bob,me,[|access(P,O,read)|]),
        pubkey(bob,rsa:3:c1ebab5d).
```

**Top-down to Bottom-up Rewrite:** Most practical access control languages, including Binder, utilize a top-down (or backward-chaining) evaluation strategy. Specific requests are made as goals, which are then resolved against the security policies, hence minimizing the disclosure of sensitive information. This suggests that *LBTrust* needs to support top-down evaluation. One possible approach that we are exploring is converting a "pull" request in the body of a rule into two "pushes". The following meta-rules express this automatic conversion:

```
pull0: says(me,X,[|request(R).|]) <-
        active([| A <-says(X,me,R), A*. |]),
        X!=me.
pull1: says(me,X,R) <-
        says(X,me,[|request(R).|]).
```

Rule pull0 matches any rule R that has `says` in the body, and dispatches a request to X. Rule pull1 responds to a request with the desired data.

### 5.2 Secure Network Datalog

*SeNDlog* [4, 19] is a unified declarative language for network specifications and security policies, which combines the Network Datalog language used in declarative networking[9] with Binder. Similar to Binder, *SeNDlog* allows different principals or contexts to communicate via import and export of tuples. To differentiate from local predicates, an *import predicate* from a principal N is quoted using "N says", whereas an *export predicate* of the form "p@X" in a rule head indicates the predicate p is exported to the principal X from the context where it is derived. We illustrate *SeNDlog* using the following two rules, s1-s2, that compute all pairs of reachable nodes in a network:

```
At S:
 s1: reachable(S,D) :- neighbor(S,D).
 s2: reachable(Z,D)@Z :- neighbor(S,Z),
                        W says reachable(S,D).
```

Rule s1-s2 are executed in the context of node S. Rule s1 takes `neighbor` tuples as input to compute one-hop `reachable` tuples. Rule s2 specifies a distributed transitive closure computation, expressing that "if Z is a neighbor of S, and S can reach D, then Z can also reach D." Unlike an ordinary transitive closure computation, the above *SeNDlog* program is

*authenticated* (via the use of "`says`") and *distributed* via the use of import and export predicates. By modifying this simple example, one can easily construct more complex secure networking protocols, such as an authenticated path-vector protocol.

Given the `says` predicate described in Section 4, the *LBTrust* equivalent of the above *SeNDlog* rules is as follows:

```
lc1: neighbor(S,D) -> prin(S), prin(D).
lc2: reachable(S,D) -> prin(S), prin(D).

ls1: reachable(me,D) <- neighbor(me,D).
ls2: says(me,Z,[|reachable(Z,D).|]) <-
        neighbor(me,Z),
        says(W,me,[|reachable(me,D).|]).
```

Using LogicBlox's support for distribution described in Section 3.5, one can customize the locations of principals by using the following rules:

```
ld1: loc(P,N) -> prin(P), node(N).
ld2: predNode(export[P],N) <- loc(P,N).
```

Rule `ld1` defines the predicate `loc` which indicates the mapping between principals and physical nodes. And rule `ld2` utilizes the built-in system predicate `predNode` to assign the physical locations of the `export` predicate according to the `loc` predicate, indicating the physical destinations of communication between principals. Users can easily enforce various distribution plans by modifying the `loc` table. Note that distribution is not required for the `neighbor` and `reachable` tables since they are only used locally at each node (and hence no partitioning is strictly necessary).

## 6. PRELIMINARY EVALUATION

A prototype of the *LBTrust* system is currently being developed. *LBTrust* leverages the LogicBlox runtime system, which has been enhanced to support meta-programmability, meta-constraints and cryptographic capabilities. We are still in the process of adding partitioning and distribution capabilities to LogicBlox, which will be useful for enforcing security policies in a distributed fashion.

We provide a preliminary evaluation of the current *LBTrust* prototype. Our evaluation consists of a micro benchmark, in which two principals `alice` and `bob` each execute a Binder rule. Together, the two principals export and import authenticated facts from each other's context via the `says` construct. In addition to plaintext transfer (which requires no signature), we measure the performance overhead of two authentication schemes using the customizable authentication rules described in Section 4: (1) *RSA* which utilizes 1024-bit RSA signatures given an input fact, and (2) *HMAC* (keyed-Hash Message Authentication Code) that generates a 160-bit SHA-1 cryptographic hash from the input fact and a secret key.

Figure 2 shows the query execution time for each experimental run, where each run consists of an increasing number of messages being exported and imported between `alice` and `bob` during rule execution. Each message results in a signature generation and verification when exported and imported. The experiment was carried out on quad-core machines with Intel Xeon 2.33GHz CPUs and 4GB RAM running Fedora Core 6 with kernel version 2.6.20. Our results validate that *LBTrust* can support various authentication schemes in a similar fashion as Binder. Moreover, *LBTrust* achieves an expected linear increase in execution time as the number of messages transferred increases. The observed performance differences are also as expected: compared to *Plaintext* where no authentication is used, *HMAC* incurs a slight increase in execution time. *RSA* is the most expensive due to the use of public key cryptography.

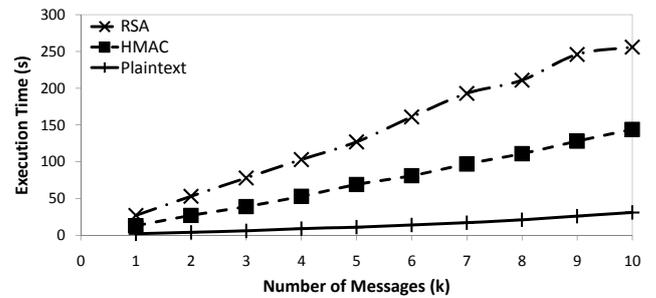

**Figure 2: Execution Time over Number of Messages**

## 7. CONCLUSION

In this paper, we present *LBTrust*, a unified declarative system for reconfigurable trust management, where various security constructs can be customized and composed in a declarative fashion.

Our work is proceeding along several fronts. Our immediate task involves prototyping a variety of recently proposed logic-based trust management systems in *LBTrust*, and utilizing our system as a basis for comparison and analyzing the security properties of these systems. We conjecture that having a unifying declarative platform will facilitate cross-language analysis and performance comparisons, and enable us to rapidly prototype new systems with novel security properties.

Second, traditional database optimizations such as magic-sets [6] can potentially bridge the top-down evaluation approach used in access control, versus the typical bottom-up continuous evaluation of network protocols. We hope to explore the use of a query optimizer to adaptively choose between two different approaches.

Third, we are currently adding provenance support to *LBTrust*. In addition to reasoning about delegation and chains of trust, provenance is useful for analyzing derivations of security policies, runtime verification, and dynamic type checking.

## 9. DEMONSTRATION PROPOSAL

Our demonstration consists of a multi-user file system with access control capabilities implemented using a combination of Binder's *authentication* and D1LP's *delegation* constructs, enabled by *LBTrust*. To facilitate our presentation, we will utilize a visualization tool used in LogicBlox to display a table of the values of various predicates and rules stored at each principal, as well as a graphical visualizer that illustrates communication between principals. Our implementation of the partitioning and distribution features mentioned in this paper is still ongoing. Our demonstration will instead be constrained to a setting where the file system runs on a single on-site laptop. Multiple principals will share a single *LBTrust* workspace under the control of a single *LBTrust* instance. While the setting is simplified, the demonstration showcases two important aspects of *LBTrust*: its meta-programmability and reconfigurable capabilities for enforcing security policies declaratively.

In our file system, each principal maintains a number of files owned by other principals. We assume the files are not necessarily stored on their owners' machines. This assumption typically holds in a distributed setting where principals may reside on different physical machines. Although our demo will be restricted to a single machine, we will emulate a distributed system via horizontal partitioning of permissions and access control policy facts among different principals. Since our special partitioning syntax is not yet implemented, we will represent a predicate's partition by adding an additional argument. Meta-constraints will be used to enforce the partition, and protect application and security predicates from unauthorized modification.

When a file access request is received, the owner of the file has the authority to grant or reject the request based on his access control policy stored in his `permission` table. For simplicity, we will present our example here based on a pre-defined `permission` table. In practice, the table itself can also be defined in terms of a number of other rules.

Figure 3(a) shows a typical workflow diagram of a principal requesting read access to a file, where principal `Requester` asks for a file `F` owned by principal `FileOwner`. `Requester` first sends a request to `FileStore`, namely the principal where `F` is stored. `FileStore` then refers to `FileOwner` to check whether `Requester` is permitted to read the file, and `FileOwner` will make the decision according to its local `permission` table. As long as the read access is granted, `FileStore` returns the file content to `Requester` and finishes the read operation.

**Example Rules** Using the Binder language described in Section 5.1, the following rules implement access control policies in a file system:

```
f1: file(F) -> .
f2: filename(F,S) -> file(F), string(S).
f3: filedata(F,S) -> file(F), string(S).
f4: fileowner(F,O) -> file(F), prin(O).
f5: filestore(F,P) -> file(F), prin(P).
f6: file(F) -> filename(F,_), filedata(F,_),
     fileowner(F,_), filestore(F,_).

m1: message(M) -> .
m2: message:id(M,N) -> message(M), int[64](N).
m3: message:fname(M,F) -> message(M), string(F).
m4: message:data(M,D) -> message(M), string(D).
m5: request(R) -> message(R).
m6: response(R) -> message(R).

dfs1: permission(P,X,F,M) -> prin(P), prin(X),
         file(F), mode(M).
dfs2: says(me,U,[|
    response(R), message:fname(R,S) <- A*. |]),
    fileName(F,S), fileowner(F,O) ->
        says(O,me,[| permission(O,U,F,read) |]).
```

In the above *LBTrust* program, we present an extension to a user-defined type `file` in rules `f1-f6`, and enumerate the types of messages exchanged when reading a file in `m1-m6`. Rules `dfs1` and `dfs2` enforce that principals only respond to authorized read requests. Note that we omit rules used to initiate and respond to read requests, as well as rules used to query the file owner for permissions. Write access control is implemented using a similar approach. We consider the problem of locating the physical machine where each file resides to be an orthogonal search problem.

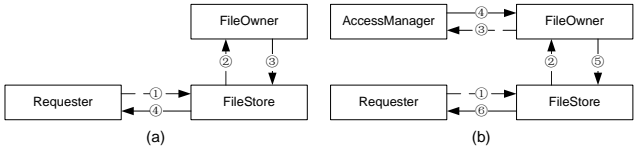

Figure 3: **Workflow of Read Access in File System**

**Access Control with Delegation** As an enhancement to the above program, one can incorporate the notion of delegation of access control in a distributed file system. We make a revision to the previous example as follows: as the `FileOwner` may have limited knowledge on the trustworthiness of other principals, it further refers to the `AccessManager` that is trusted to decide whether a principal is permitted to perform operations on a specific file. As shown in Figure 3(b), the `FileOwner` *delegates* the authority of making access control decisions to a trusted `AccessManager`.

Specifying delegation demands a security construct used in prior work on delegation logic [15]. In our demonstration, we use a slight modification of the user-composed security predicate `delegates` presented in Section 4.2. The file owner would delegate his `permission` predicate to `accessMgr` using the following rule:

```
delegates(me,accessMgr,[|permission(me,_,F,_).|]) <-
    fileowner(F,me).
```

As our demonstration is confined to a single workspace, we do not suffer from the drawback described in Section 4.2, and can easily enforce delegation depth constraints. We will show delegations with depth restrictions (`AccessManager` is not allowed to further delegate `permission`), and threshold restrictions (`FileOwner` permits a file access request only if the permission is confirmed by more than three `AccessManagers`).